\documentclass[10pt,aps,prr,final,twocolumn,amsmath,amssymb,superscriptaddress,longbibliography]{revtex4-1}
\usepackage[pdftex, colorlinks=true, linkcolor=blue, citecolor=blue, urlcolor=blue]{hyperref}
\usepackage{graphicx}
\usepackage{aurical}
\usepackage[T1]{fontenc}
\usepackage[version=3]{mhchem}
\usepackage{siunitx}
\usepackage{braket}
\graphicspath{{./figures/}}
\DeclareGraphicsExtensions{.png,.pdf}

\newcommand{\una}{Universit\'{e} de Nouakchott, Facult\'{e} des Sciences et Techniques, D\'{e}partement de Physique, Avenue du Roi Fai\c{c}al, 2373, Nouakchott, Mauritania}
	
\begin{document}
\title{Magnetic dynamics driven integer and fractional high harmonic generation arising from highly nonlinear instantaneous energy levels}
\author{O. Ly}
\email{ousmanebouneoumar@gmail.com}
\affiliation{\una}

\begin{abstract}
We demonstrate that the strongly nonlinear regime of the magnetic dynamics driven high harmonic generation (HHG) 
can be captured through an adiabatic treatment. {The underlying} instantaneous energy dispersion reveal a highly nonlinear term that mirrors the high-frequency excitations present in non-equilibrium transport quantities. { We stress out that the instantaneous energy levels dynamics carry much information on HHG and thereby provide a simplified understanding of the effect. Further, we exploit the present approach to}
predict the emergence of tunable integer and fractional high harmonics in the presence of a time-modulated spin-orbit interaction.  
{Finally, we apply the approach to the case of the light driven HHG to figure out differently emerging non-linearities.}
Our findings initiate {a new direction for harnessing HHG in both laser and precession driven ultrafast carrier dynamics.} 
\end{abstract}
	
\maketitle
\section{Introduction}
When a gas \cite{Ferray1988,  McPherson1987, Popmintchev2012}, a liquid \cite{Luu2018, Mondal2023} or a solid state system \cite{Schubert2014, Ghimire2019, Ghimire2011, Wu2015} is {excited by a laser beam} 
operating at a frequency $\omega$, an emission at higher frequencies {can be} obtained. This phenomenon is known as higher harmonic generation (HHG).
In the context of quantum optics, this generation is often interpreted within the so-called three step model, where the emission results from a loss of kinetic energy when an electron is accelerated and subsequently decelerated by the laser electric field modulation. This effect has been widely investigated in the past three decades, leading to the well established field of atto-physics \cite{Corkum2007, Ghimire2014, Krausz2016, Krausz}.
	
It was only very recently that we proposed the use of adiabatic magnetic precession instead of light to induce similar {HHG} in a spintronic system as a result of the interplay between magnetic dynamics and spin-orbit coupling \cite{Ly2022} or non-collinear real space magnetic textures \cite{Ly2023}.

In contrast to conventional pumping (in the absence of spin-orbit coupling) from ferromagnetic \cite{Tserkovnyak2002, Tserkovnyak2002b, Tserkovnyak2005} or antiferromagnetic  \cite{Cheng2014, Vaidya2020} orders, where all a.c currents oscillate with the same frequency as the drive, our proposal suggested the emergence of ultrafast carrier dynamics in the presence of spin-flip scattering terms.
Nonetheless, a complete consistent understanding of the magnetically induced generation in all relevant regimes has been so far elusive.
In the present letter, we undertake the task of studying the elementary key aspects that allows us {to deepen our understanding of this effect} as well as the role played by the underlying pertinent parameters. 
To this end, we consider a very simple system (a ferromagnetic Rashba chain). {However, despite its apparent simplicity our model contains all the} ingredients necessary for the emergence of high order harmonics.
	
{In our study, we will focus on the magnetic system} from a pure analytical {perspective} in order to reveal interesting features related to the time dependent spin resolved band structure. 
{We find that HHG is a property of the instantaneous energy levels developing in the presence of a precessing magnetic order which acts as a drive.
Further, we demonstrate that the nonlinear dynamics in the bands appear similarly in the exactly numerically computed carrier dynamics. {Our findings suggest} that both instantaneous bands and transport quantities {scales up similarly in terms} of the main parameters governing the generation effect. This suggests an appealing perspective for harnessing HHG in magnetic systems. And could be presumably generalized to study a broad spectrum of driven condensed matter systems.} {Furthermore, we utilise the present formalism to predict unforeseen generation mechanisms in the magnetic Rashba system. We find that when the Rashba field is time modulated in the presence of a precessing magnetic order, tunable integer and non-integer HHG can emerge.}
	
\begin{figure}
	\includegraphics[width=0.5\textwidth]{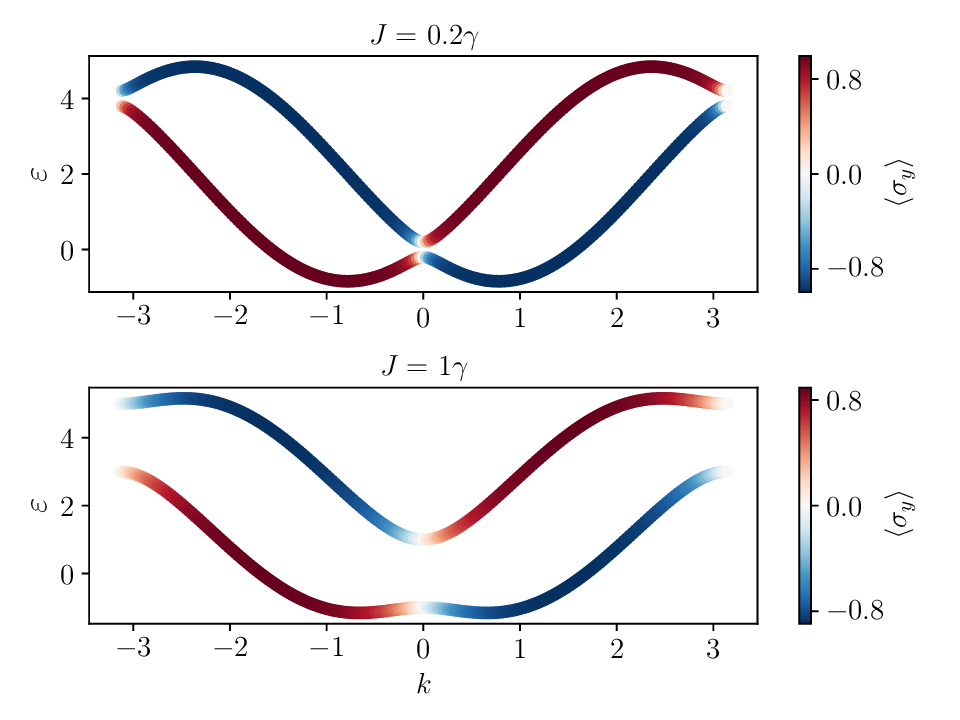}
\caption{The band dispersion for two different values of the s-d exchange coupling $J$. The precession angle is set at $\theta=\pi/2$ and the Rashba constant is $\alpha=\gamma$. The color axis displays the spin expectation value along the transverse direction $y$.}
			\label{fig1}
\end{figure}

\section{Analytical description of the time dependent magnetic Rashba chain}
The study of quantum transport in {systems with non-negligible spin-orbit coupling is a familiar problem in mesoscopic physics.} 
{For instance,} it has been investigated in the context of designing spin filters in quasi one-dimensional structures \cite{Streda2003, Birkholz2008}.  
{In these works, the authors considered an arbitrary orientation for the applied Zeeman field. Although this would be similar to a field whose orientation changes in time, the behaviour of the related transport signals in the time domain was {not investigated therein}.}
{In the present case}, we consider a time dependent magnetization which can be described in a similar fashion as a Zeeman interaction. 
	
{Since we are interested in identifying the basic concepts behind the emergence of non-linear dynamics in the spin transport context we consider a minimal setup consisting of}
	a one dimensional ferromagnet, with s-d exchange coupling $J$, in the presence of a Rashba like spin-orbit interaction with a constant strength  $\alpha$. 
	
	The corresponding time dependent tight-binding Hamiltonian is defined as follows:
	\begin{equation}
		\mathcal{H} = \mathcal{H}_0(t) + \mathcal{H}_{\rm R},
	\end{equation}
	{with}
	\begin{equation}
		\mathcal{H} _0(t)=  
        J\sum_{i}\hat{c}^{\dagger}_i(\mathbf{\sigma}\cdot\mathbf{m}(t))\hat{c}_i-\gamma\sum_{\langle i, j \rangle}(\hat{c}^{\dagger}_i\hat{c}_j + \rm{H.c}),
	\end{equation}
	where $ \rm{H.c}$ is the hermitian conjugate. The time dependent magnetization  at an instant $t$ is given as $\mathbf{m}(t)=(\sin \theta \cos {\omega t}, \sin \theta \sin {\omega t}, \cos{\theta})$, and is therefore assumed to  be precessing at an angle $\theta$ and frequency $\omega$. 
	The entities $\hat{c}^{\dagger}_i$ and $\hat{c}_i$ represent respectively, the creation and annihilation operators at site $i$. {We note that $\sigma$ is the usual three dimensional vector of Pauli matrices. The pre-factor $\gamma$ stands for the tight binding hopping parameter.}
	
	We also include a one dimensional Rashba term, which is given by
	\begin{equation}
		\mathcal{H} _{\rm R}= \alpha \sum_{j}(\hat{c}^{\dagger}_j i\mathbf{\sigma}_y\hat{c}_{j+1} + \rm{H.c}).
	\end{equation}
	 
	{We will be assuming an adiabatic approximation. - that is a slow precession of the magnetic drive. } 
    {The corresponding instantaneous bands}
    are obtained as \cite{Birkholz2008}
	\begin{equation}\label{bands}
		\varepsilon_k^{\pm}(t) =  - 2\gamma\cos k \pm 2  \mathrm{sgn}(k - k_0(t))\sqrt{D_k(t)},
	\end{equation}
	where 
	$$k_0(t)=\arcsin (\frac{J}{2 \alpha}\sin \theta \sin{\omega t})$$ 
	and 
	\begin{equation}
	\label{eq:dkt}
	D_k(t)=\alpha^2\sin^2 k + J^2/4 - J\alpha\sin k \sin \theta \sin {\omega t}.
	\end{equation}
	The superscript $\pm$ accounts for the two energy branches with opposite spin expectation values along the transverse direction $y$.
	
	Further, the spin expectation values can be obtained as :
	\begin{equation}\label{eq:sxy}
		\langle \sigma_x - i\sigma_y \rangle_k^{\pm} = 2\frac{C_k^{\pm}}{1+\left |C_k^{\pm}\right |^2},
	\end{equation}
	and 
	\begin{equation}\label{eq:sz}
		\langle \sigma_z \rangle_k^{\pm} = \frac{-1 + \left |C_k^{\pm}\right |^2}{1+\left |C_k^{\pm}\right |^2},
	\end{equation}
	where we have introduced the momentum dependent coefficient 
	\begin{equation}
		C_k^{\pm}=\frac{-i\alpha\sin k-\frac{J}{2} e^{-i\omega t}\sin \theta}{\frac{J}{2}\cos \theta  \mp {\rm {sgn}}(k-k_0(t))\sqrt{D_k(t)}}.
	\end{equation}

	Although, these results were derived in ref. \cite{Birkholz2008}, the authors did not consider so far the corresponding time domain transport. 
	In the present case, our focus will be mainly on the carrier dynamics underlying the magnetic Rashba chain.
	
	\begin{figure}
	\includegraphics[width=0.48\textwidth]{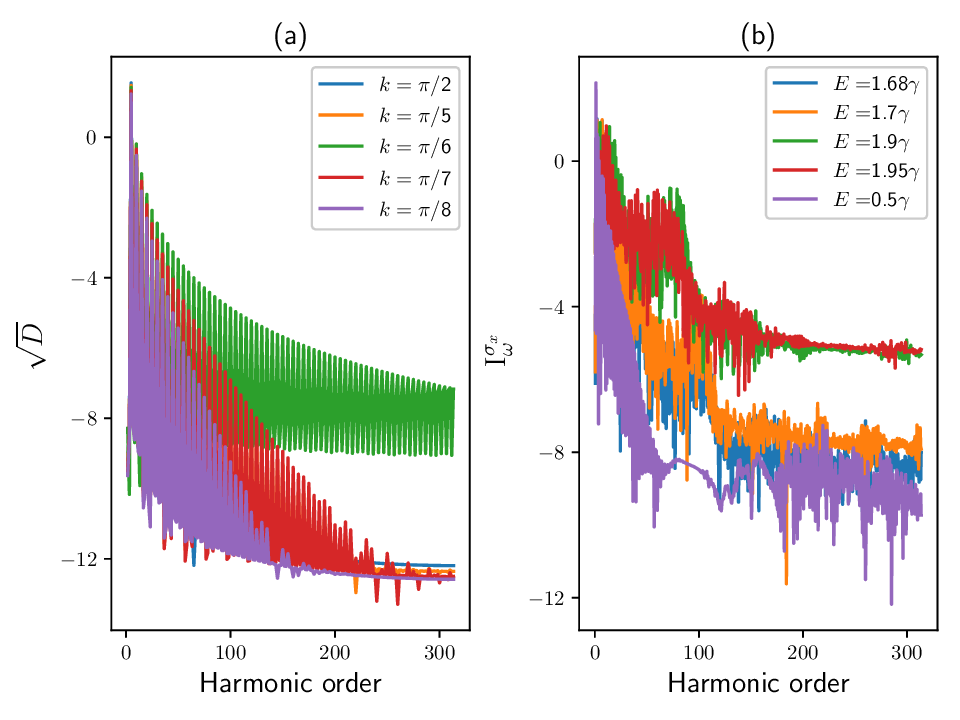}
	\caption{The Fourier transform of the quantity $\sqrt{D}$ at different momenta is shown in panel (a). Panel (b) displays the frequency domain response of the spin current along $x$ for different energy values obtained from time dependent transport calculations via Tkwant \cite{tkwant}. The parameters of the considered system are : $J=\alpha=\gamma$, $\theta=\pi/2$ and $\omega=0.01\gamma/\hbar$. 
    }
	\label{fig2}
    \end{figure}

\section{Integer harmonics in carrier pumping arising from highly nonlinear energy bands} 
To comprehend the underlying factors contributing to the emergence of high-frequency excitations in the magnetic Rashba system, we employ the above analytical description of the energy levels within an adiabatic treatment. 
Our analysis begins by examining the energy levels at a fixed orientation of the magnetization vector.
In Fig. \ref{fig1}, the spin-resolved band structure is depicted for two values of the s-d exchange constant at a fixed Rashba strength ($\alpha=1\gamma$). The time-reversal-breaking dynamical magnetic moment induces a pseudo energy gap (minimum splitting of the two levels) \cite{Streda2003}, which becomes momentum and time-dependent as the magnetization precesses. 
It turns out that the dynamics of the energy levels is an essential precursor for the ultrafast dynamics underlying the quantum transport properties of the magnetic system.
Starting from the analytical description above, one first notice the term $\sqrt{D}$ which exhibits a highly nonlinear nature.  
This non-linearity extends to the wave-functions and, consequently, the spin expectation values (Eq. \eqref{eq:sxy} and \eqref{eq:sz}) through the coefficients $C_k^{\pm}$. 
{Even in a more general case, non-linearities in instantaneous energy bands will propagate to the wavefunctions and thereby to the underlying transport quantities. This makes our argument rather general. Therefore, we expect these non-linearities to manifest similarly in the quantum transport response of a more general system. 
}

To elucidate the impact of these highly nonlinear energy levels on the non-equilibrium transport properties of the system, we compare the dynamical term $\sqrt{D}$ with relevant transport quantities.
In Fig. \ref{fig2}, the Fourier transform of $\sqrt{D}$ is compared with the spin current computed at different energies. 
Without loss of generality, we consider the spin current polarized along the $x$ direction, $I^{\sigma_x}$. In fact the HHG is present in all spin currents in addition to the charge.
These currents are obtained 
using the non-equilibrium quantum transport package Tkwant \cite{tkwant}. 
{To compute the current flowing toward a given lead, the underlying scattering wavefunction at time $t=0$ is evolved according to the time dependent Schr{\"o}dinger equation. The corresponding spin current polarized along the direction $j$ at an ulterior time $t$ from a site $i$ in the chain to site $l$ belonging to the lead interface, is then obtained as 
$I^{\sigma_j}={\Im}\left[\psi_{i}^{\dagger}(t){\mathcal{H}_{il}(t) \sigma_j}\psi_l(t)\right]$,
where $\mathcal{H}_{il}$ is the Hamiltonian matrix element between the two sites and $\psi_i(t)$ is the scattering wavefunction at site $i$ and time $t$. For the charge current $\sigma_j$ should be replaced by the unit matrix.}

Notably, similar high harmonic features are present in both the current and the dynamical factor. 
Accordingly, one can initially state that the transport quantities can be seen as functionals of $\sqrt{D}$. 

{ To get an initial analytical insight on this correlation we consider the charge current computed within the adiabatic approximation. We note that in the presence of SOC, the adiabatic approximation is not applicable at the crossing point. However, in the presence of a ferromagnetic order such as in the present case, the degeneracy is lifted. And the adiabatic approximation is applicable.}
{
The corresponding adiabatic current is constituted of two contributions:

(i) The band velocity which is simply the derivative of $\varepsilon$ with respect to momentum. This gives simply a term $\propto 1/\sqrt{D}$.

{ 
(ii) The Berry curvature contribution. Which can be obtained for a given band $s$ with eigenstate $\ket{\psi_k^s}$, as 
$\Omega_s=2\Im \braket{\partial_k {\psi_k^s}|\partial_t {\psi_k^s}}$ \cite{Xiao2010}.
The underlying eigenstate reads $\ket{\psi_k^s}=1/\sqrt{1+|C_k^s|^2}
\begin{pmatrix} C_k^s \\  1 \end{pmatrix}
$ \cite{Birkholz2008}.
Defining 
${\tilde{C_k^s}}=C_k^s/\sqrt{1+|C_k^s|^2}$, the curvature reduces to
$\Omega_s=2\Im\left[ \partial_k {\tilde{C}}_k^{s\dagger} \partial_t {\tilde{C}}_k^{s}\right]$.  
}

Computing the underlying expression is cumbersome indeed, but leads straightforwardly to an explicit dependence on $\sqrt{D}$ as $C_k^s$ is a simple functional of $\sqrt{D}$. A summation of these contributions over the momentum space should be of course performed to obtain the full response. 
It shall be understood that all nonlinear forms or functionals of ${D}$ should display similar nonlinear features as $\sqrt{D}$. And so is the adiabatic current. 
}
{
We shall emphasize that our calculations are performed at $\theta=\pi/2$. However, our results remain valid in general at arbitrary angles as demonstrated in our previous studies \cite{Ly2022, Ly2023c}. We note that in a two terminal geometry, the maximal excitation of harmonics corresponds to $\theta=\pi/2$.
Notwithstanding, we do not expect the adiabatic results obtained for an infinite (uniform) system to lead to identical results as our hybrid geometry, which consists of a magnetic system attached to non-magnetic leads. Nonetheless, the overall dependence of the currents on $\sqrt{D}$ is expected to manifest similarly.
Further, it is of importance to mention that conventional spin pumping devices usually operate at low precession angles. However, the strong emission at large angles can be triggered in current driven nano-oscillator devices \cite{Kiselev2003, Liu2012}, where an injected spin current causes the magnetic order to precess with a large angle. 
}

\begin{figure}[htbp]
\includegraphics[width=.45\textwidth]{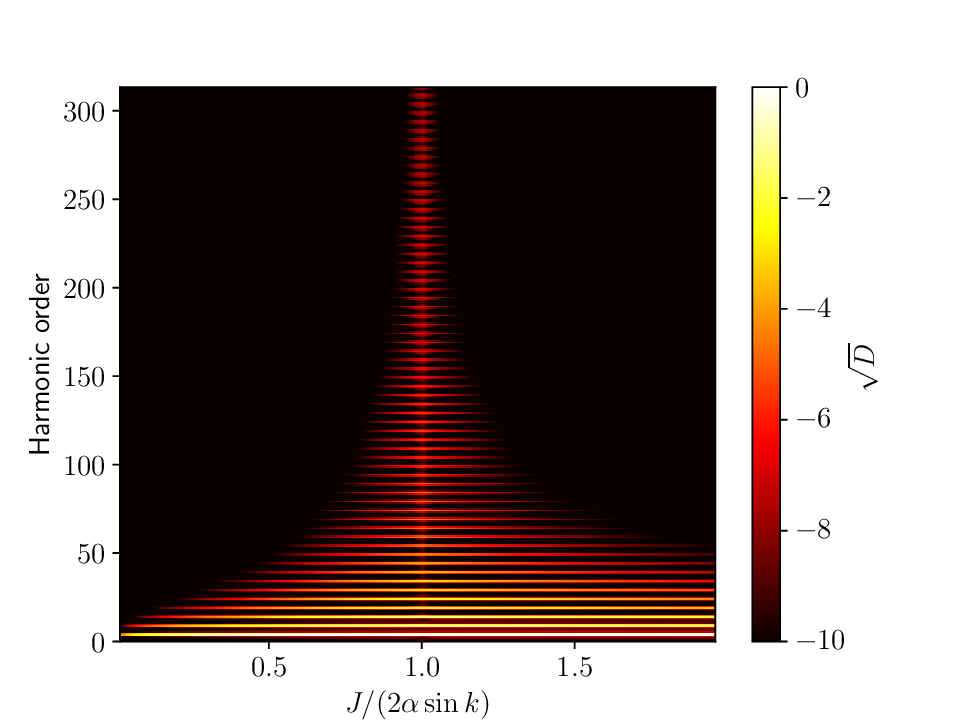}
\caption{The Fourier domaine representation of $\sqrt{D}$ is shown as a function of the s-d exchange coupling $J$ at $\theta=\pi/2$.}
\label{fig3}
\end{figure}

An important aspect related to the magnetic dynamics driven HHG effect concerns the way the transport signals scale up with different parameters of the problem. Hence, 
to further consolidate the statement above, we assess the scaling of $\sqrt{D}$ in terms of $J$ and $\alpha$. 
For the scaling of the spin currents in terms of these parameters, we base on our previous numerical simulations \cite{Ly2022, Ly2023c} of the effect where a particular resonance feature corresponding to a maximum emission was obtained when $J/k\alpha$ is appropriately tuned. 
{In a realistic material, $J$ can range from $0.1$eV to few eVs \cite{Cooper1967, Xiao2008}. Therefore, a strong SOC is required for the matching condition in the high s-d exchange coupling limit. This can be achieved in Bi/Ag alloys or Bismuth based topological materials, where $\alpha$ can be as high as few eVs \cite{Ast2007, Zhang2009}.}
Of major interest, is the resonance emission (ultrahigh harmonic regime), where the generation is maximized at a given {ratio} between $\alpha k$ and $J$.
Yet, we concentrate on how $\sqrt{D}$ (and consequently $\varepsilon$) behaves as $J/\alpha$ is varied. 
In Fig. \ref{fig3}, the same resonance behaviour as in our previous simulations is observed. 
Since the previous results were constructed from numerical observations, the resonance condition was not accurately reflected. 
There, we have had simply guessed that a resonance emission corresponding to the ultrahigh harmonic regime might occur when the ferromagnetic gap matches the Rashba splitting - that is when $J/2k\alpha=1$. 
Now observing the analytical form of $\sqrt{D}$, we deduce that the resonance condition should rather correspond to ${J}=2\alpha \sin k$. 
{Though, the initial condition remains valid in the low momentum limit.} 
Our computation of the Fourier domain of the dynamical term essentially captures this resonance feature, where a dramatic enhancement of the emission bandwidth is obtained at this matching condition.
This observation further underscores a pronounced correlation between carrier currents and instantaneous band dynamics. And therefore confirms our initial statement that the ultrafast carrier dynamics is typically coming from the highly nonlinear nature of the instantaneous energy dispersion. 
{This is consistent with a recent suggestion of the prevalence of the adiabatic energy levels picture in HHG \cite{Neufeld2022}.}
Next, we will exploit this approach to predict new generation scenarios in a tailored magnetic Rashba system. 

\section{Tunable integer and fractional high harmonics in the presence of a time dependent Rashba field}
An essential facet of our approach lies in its ability to predict interesting features of time-domain transport responses through a straightforward analysis of instantaneous energy levels. To illustrate this capability, we investigate a scenario involving dynamic spin-orbit coupling combined with magnetic precession. The tunability of the Rashba interaction via external fields makes it particularly attractive for spintronic applications, as demonstrated in the design of a dynamical spin-orbit-based spin transistor in a non-magnetic system \cite{Gursoy2023}, where integer harmonics were observed in the emitted spin currents. In our case, we integrate both dynamic Rashba coupling and magnetic precession.
When both Rashba and magnetic interactions are time-dependent, one anticipates broadband components in the time-domain signal due to the interplay between these dynamics. However, the resulting generation should mainly depend on the relative ratio between the underlying frequencies. 

\begin{figure*}
\includegraphics[width=0.8\textwidth]{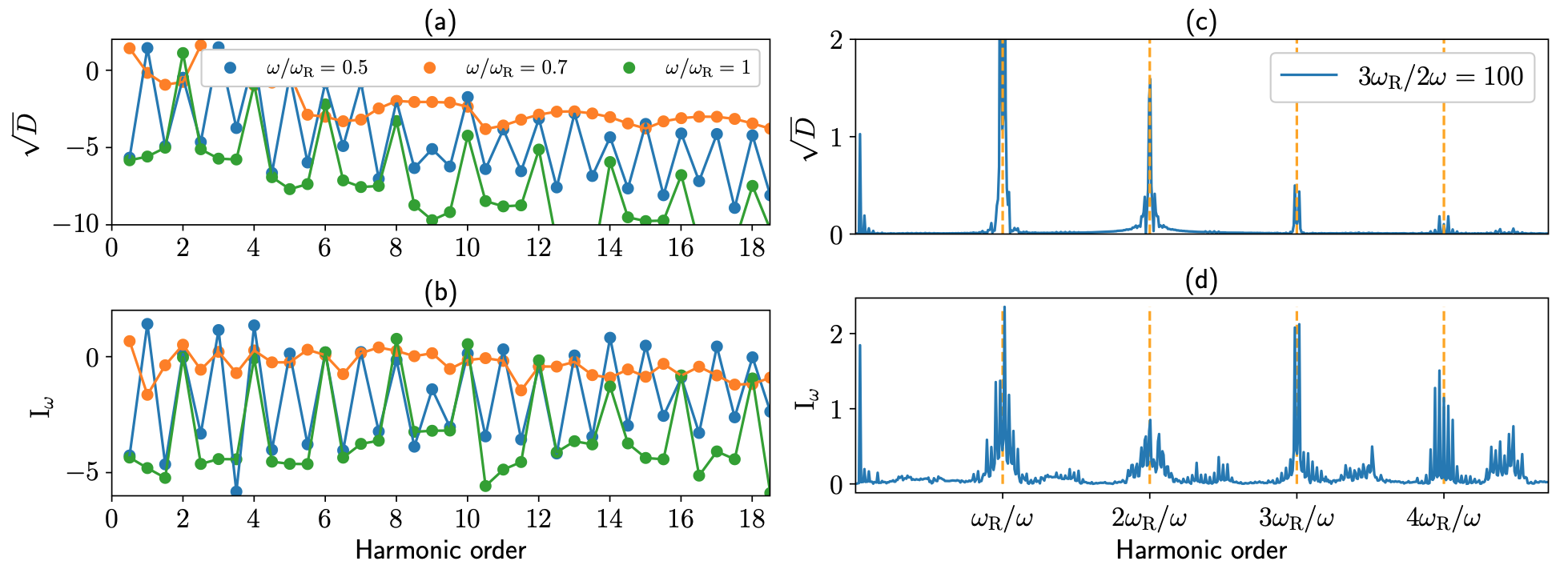}
\caption{The Fourier transform of $\sqrt{D}$ is represented in panel (a). Panel (b) displays the frequency domain representation of the charge current ${\rm{I}}_{\omega}$ obtained form the time dependent quantum transport package \cite{tkwant}. 
In panel (c) and (d) $\sqrt{D}$ and ${\rm{I}}_{\omega}$ are displayed respectively for a large value of the Rashba frequency $\omega_{\rm R}=66.66\omega$.
The cone angle of the dynamics is set at $\theta=\pi/2$. The Rashba strength is modulated in time as $\tilde{\alpha}(t)=\alpha\sin \omega_{\rm R} t$. We took $J=1\gamma$. The quantity $\sqrt{D}$ is evaluated at a fixed momentum value $k=\pi/6$. The charge current is computed at an energy value $\varepsilon=-1.9\gamma$.}
\label{dyna}
\end{figure*}

For our analysis, we consider the time-modulated Rashba strength in the form $\tilde{\alpha}(t) = \alpha\sin \omega_{\rm R} t$, with $\omega_{\rm R}$ representing the modulation frequency induced by a time-dependent external electric field oscillating at the same frequency. This leads to modified instantaneous energy bands, where the dynamical term in Eq. \eqref{eq:dkt} becomes $D_k(t)=\alpha^2\sin^2\omega_{\rm R}t\sin^2 k + J^2/4 - J\alpha\sin k \sin \theta \sin {\omega t} \sin\omega_{\rm R}$, revealing the potential for exciting different frequencies when $\omega/\omega_{\rm R}$ is appropriately tuned.
We anticipate three regimes of HHG based on our initial insight: (i) when $\omega_{\rm R}=2\omega$, all integer harmonics are expected; (ii) when $\omega_{\rm R}=\omega$, only even harmonics are excited; (iii) when $\omega_{\rm R}/\omega$ is not an integer, intermediate frequencies can be excited.
Numerical computations of the charge current emitted from the dynamical Rashba chain validate our predictions. Fig. \ref{dyna} illustrates the frequency spectra of $\sqrt{D}$ compared with the numerically obtained charge current at a fixed energy. The observed emergence of the same set of frequencies in both quantities aligns precisely with our initial analysis of the instantaneous energy levels of Eq. \eqref{bands}. In cases where $\omega/\omega_{\rm R}$ is non-integer (the orange curve in Fig. \ref{dyna}), the emission is broadband, displaying a continuous range of frequencies with comparable amplitudes. In the other two cases, where $\omega_{\rm R}$ is an integer multiple of $\omega$, two sets of harmonics are observed, with the blue (green) curves representing cases where both even and odd integers (even integers) are emitted, respectively.
We further explore the scenario where $\omega_{\rm R}$ is much larger than $\omega$, predicting that larger frequency excitations at $n\omega_{\rm R}/\omega$ dominate the carrier responses. Panel (d) of Fig. \ref{dyna} confirms these predictions, showing the corresponding charge current with similar frequency domain features as in the dynamical factor.

It's noteworthy that in the present case where inversion symmetry is dynamically broken, the pumped currents exhibit substantial DC components (not shown), in contrast to the case where a constant Rashba interaction is considered. In the latter scenario the DC component of the charge current is found to be relatively small \cite{Ly2022}, even in the case of an asymmetric geometry such as an antiferromagnet.

\section{Discussion and Conclusion}
The proposed methodology explores the potential of harnessing magnetic dynamics driven HHG through an analysis of instantaneous energy levels. Our approach 
simplifies the study of HHG, making it accessible even at an elementary level. This accessibility allows for unprecedented control over the relevant parameter space, facilitating the optimization of underlying carrier emission.


It is worth noting that our approach {might be applicable for} 
understanding similar generation mechanisms, such as those observed in recent studies on HHG in correlated electronic systems \cite{Imai2020}, valley pumping setups \cite{Hadadi2023}, systems with dynamical Rashba effects in nonmagnetic environments \cite{Gursoy2023}, generalized non-collinear textures induced HHG \cite{Ly2023} { and carrier-envelope phase engineered fractional emission in the presence of topological insulators \cite{Schmid2021}}. Additionally, our prediction of tunable broadband emission from time-modulated Rashba interactions presents further exciting possibilities for engineering THz emission in spintronic devices.

We shall emphasize that carrier pumping in spintronics is typically described as an adiabatic process as it usually involves relatively low driving frequencies. Therefore, our approach remains realistic within this context. However, exploring the approach in the high-frequency limit could provide further interesting theoretical insights.

{So far, we have focused on the case where HHG is induced by magnetic precession. Therefore, we find it useful to analyse the case where the generation is rather induced by a laser beam. To this end, we consider a laser field with a driving frequency $\omega_L$ in our analytical model (without magnetic drive, $J=0$). To account for the effect of the laser excitation, we perform the standard Peirl's substitution in the energy levels, by simply shifting the momentum $k$ by the underlying uni-directional vector potential, which is given as $A(t)=a\cos(\omega_L t)$. Where $a$ stands for the amplitude of the driving potential. In this case, the corresponding dynamical term becomes $\propto\sin(k+a e/\hbar \cos(\omega_L t))$. With $e$ being the elementary charge and $\hbar$ the reduced Planck constant. Clearly, in this case the dynamical term presents a highly nonlinear nature leading to the emergence of HHG. We point out that such type of non-linearities ($\sin(k+a e/\hbar \cos(\omega_L t))$) have been reported in a light driven spin-orbit coupled system \cite{Lysne2020}. This confirms the applicability of the present approach beyond magnetic dynamics driven HHG. 
}

In summary, our study of the instantaneous bands in the magnetic Rashba chain reveals the emergence of ultrafast carrier dynamics driven by highly nonlinear instantaneous energy levels. 
First, we have numerically demonstrated a strong correlation between the unveiled highly nonlinear energy levels and the underlying exactly computed carrier dynamics. 
To confirm this relationship, we have further studied the scaling up of the relevant quantities in terms of key parameters of the problem.
In order to validate our initial insights, we have built upon the dynamics of the instantaneous levels to predict unforeseen sets of tunable integer and fractional harmonics in the case of a dynamical spin-orbit interaction. This prediction was subsequently confirmed by further exact numerical simulations.
Our approach demonstrates that simply analyzing instantaneous energy levels can unveil unique physical properties without (or prior to) resorting to demanding non-equilibrium quantum transport calculations. And thereby initiates a tantalizing perspective for studying time dependent quantum transport phenomena in condensed matter systems. 

\acknowledgments
We thank M. Vogl for useful comments, suggestions and valuable discussions and inputs. We also thank A. Abbout, H. M. Abdullah and M. A. Ami for useful discussions. 
\bibliography{refs}

\begin{thebibliography}{37}%
\makeatletter
\providecommand \@ifxundefined [1]{%
 \@ifx{#1\undefined}
}%
\providecommand \@ifnum [1]{%
 \ifnum #1\expandafter \@firstoftwo
 \else \expandafter \@secondoftwo
 \fi
}%
\providecommand \@ifx [1]{%
 \ifx #1\expandafter \@firstoftwo
 \else \expandafter \@secondoftwo
 \fi
}%
\providecommand \natexlab [1]{#1}%
\providecommand \enquote  [1]{``#1''}%
\providecommand \bibnamefont  [1]{#1}%
\providecommand \bibfnamefont [1]{#1}%
\providecommand \citenamefont [1]{#1}%
\providecommand \href@noop [0]{\@secondoftwo}%
\providecommand \href [0]{\begingroup \@sanitize@url \@href}%
\providecommand \@href[1]{\@@startlink{#1}\@@href}%
\providecommand \@@href[1]{\endgroup#1\@@endlink}%
\providecommand \@sanitize@url [0]{\catcode `\\12\catcode `\$12\catcode `\&12\catcode `\#12\catcode `\^12\catcode `\_12\catcode `\%12\relax}%
\providecommand \@@startlink[1]{}%
\providecommand \@@endlink[0]{}%
\providecommand \url  [0]{\begingroup\@sanitize@url \@url }%
\providecommand \@url [1]{\endgroup\@href {#1}{\urlprefix }}%
\providecommand \urlprefix  [0]{URL }%
\providecommand \Eprint [0]{\href }%
\providecommand \doibase [0]{http://dx.doi.org/}%
\providecommand \selectlanguage [0]{\@gobble}%
\providecommand \bibinfo  [0]{\@secondoftwo}%
\providecommand \bibfield  [0]{\@secondoftwo}%
\providecommand \translation [1]{[#1]}%
\providecommand \BibitemOpen [0]{}%
\providecommand \bibitemStop [0]{}%
\providecommand \bibitemNoStop [0]{.\EOS\space}%
\providecommand \EOS [0]{\spacefactor3000\relax}%
\providecommand \BibitemShut  [1]{\csname bibitem#1\endcsname}%
\let\auto@bib@innerbib\@empty
\bibitem [{\citenamefont {Ferray}\ \emph {et~al.}(1988)\citenamefont {Ferray}, \citenamefont {L'Huillier}, \citenamefont {Li}, \citenamefont {Lompre}, \citenamefont {Mainfray},\ and\ \citenamefont {Manus}}]{Ferray1988}%
  \BibitemOpen
  \bibfield  {author} {\bibinfo {author} {\bibfnamefont {M}~\bibnamefont {Ferray}}, \bibinfo {author} {\bibfnamefont {A}~\bibnamefont {L'Huillier}}, \bibinfo {author} {\bibfnamefont {X~F}\ \bibnamefont {Li}}, \bibinfo {author} {\bibfnamefont {L~A}\ \bibnamefont {Lompre}}, \bibinfo {author} {\bibfnamefont {G}~\bibnamefont {Mainfray}}, \ and\ \bibinfo {author} {\bibfnamefont {C}~\bibnamefont {Manus}},\ }\bibfield  {title} {\enquote {\bibinfo {title} {Multiple-harmonic conversion of 1064 nm radiation in rare gases},}\ }\href {\doibase 10.1088/0953-4075/21/3/001} {\bibfield  {journal} {\bibinfo  {journal} {Journal of Physics B: Atomic, Molecular and Optical Physics}\ }\textbf {\bibinfo {volume} {21}},\ \bibinfo {pages} {L31} (\bibinfo {year} {1988})}\BibitemShut {NoStop}%
\bibitem [{\citenamefont {McPherson}\ \emph {et~al.}(1987)\citenamefont {McPherson}, \citenamefont {Gibson}, \citenamefont {Jara}, \citenamefont {Johann}, \citenamefont {Luk}, \citenamefont {McIntyre}, \citenamefont {Boyer},\ and\ \citenamefont {Rhodes}}]{McPherson1987}%
  \BibitemOpen
  \bibfield  {author} {\bibinfo {author} {\bibfnamefont {A.}~\bibnamefont {McPherson}}, \bibinfo {author} {\bibfnamefont {G.}~\bibnamefont {Gibson}}, \bibinfo {author} {\bibfnamefont {H.}~\bibnamefont {Jara}}, \bibinfo {author} {\bibfnamefont {U.}~\bibnamefont {Johann}}, \bibinfo {author} {\bibfnamefont {T.~S.}\ \bibnamefont {Luk}}, \bibinfo {author} {\bibfnamefont {I.~A.}\ \bibnamefont {McIntyre}}, \bibinfo {author} {\bibfnamefont {K.}~\bibnamefont {Boyer}}, \ and\ \bibinfo {author} {\bibfnamefont {C.~K.}\ \bibnamefont {Rhodes}},\ }\bibfield  {title} {\enquote {\bibinfo {title} {Studies of multiphoton production of vacuum-ultraviolet radiation in the rare gases},}\ }\href {\doibase 10.1364/JOSAB.4.000595} {\bibfield  {journal} {\bibinfo  {journal} {J. Opt. Soc. Am. B}\ }\textbf {\bibinfo {volume} {4}},\ \bibinfo {pages} {595--601} (\bibinfo {year} {1987})}\BibitemShut {NoStop}%
\bibitem [{\citenamefont {Popmintchev}\ \emph {et~al.}(2012)\citenamefont {Popmintchev}, \citenamefont {Chen}, \citenamefont {Popmintchev}, \citenamefont {Arpin}, \citenamefont {Brown}, \citenamefont {Ali{\v{s}}auskas}, \citenamefont {Andriukaitis}, \citenamefont {Bal{\v{c}}iunas}, \citenamefont {M{\"u}cke}, \citenamefont {Pugzlys}, \citenamefont {Baltu{\v{s}}ka}, \citenamefont {Shim}, \citenamefont {Schrauth}, \citenamefont {Gaeta}, \citenamefont {Hern{\'{a}}ndez-Garc{\'{i}}a}, \citenamefont {Plaja}, \citenamefont {Becker}, \citenamefont {Jaron-Becker}, \citenamefont {Murnane},\ and\ \citenamefont {Kapteyn}}]{Popmintchev2012}%
  \BibitemOpen
  \bibfield  {author} {\bibinfo {author} {\bibfnamefont {Tenio}\ \bibnamefont {Popmintchev}}, \bibinfo {author} {\bibfnamefont {Ming-Chang}\ \bibnamefont {Chen}}, \bibinfo {author} {\bibfnamefont {Dimitar}\ \bibnamefont {Popmintchev}}, \bibinfo {author} {\bibfnamefont {Paul}\ \bibnamefont {Arpin}}, \bibinfo {author} {\bibfnamefont {Susannah}\ \bibnamefont {Brown}}, \bibinfo {author} {\bibfnamefont {Skirmantas}\ \bibnamefont {Ali{\v{s}}auskas}}, \bibinfo {author} {\bibfnamefont {Giedrius}\ \bibnamefont {Andriukaitis}}, \bibinfo {author} {\bibfnamefont {Tadas}\ \bibnamefont {Bal{\v{c}}iunas}}, \bibinfo {author} {\bibfnamefont {Oliver~D.}\ \bibnamefont {M{\"u}cke}}, \bibinfo {author} {\bibfnamefont {Audrius}\ \bibnamefont {Pugzlys}}, \bibinfo {author} {\bibfnamefont {Andrius}\ \bibnamefont {Baltu{\v{s}}ka}}, \bibinfo {author} {\bibfnamefont {Bonggu}\ \bibnamefont {Shim}}, \bibinfo {author} {\bibfnamefont {Samuel~E.}\ \bibnamefont {Schrauth}}, \bibinfo {author} {\bibfnamefont {Alexander}\ \bibnamefont {Gaeta}},
  \bibinfo {author} {\bibfnamefont {Carlos}\ \bibnamefont {Hern{\'{a}}ndez-Garc{\'{i}}a}}, \bibinfo {author} {\bibfnamefont {Luis}\ \bibnamefont {Plaja}}, \bibinfo {author} {\bibfnamefont {Andreas}\ \bibnamefont {Becker}}, \bibinfo {author} {\bibfnamefont {Agnieszka}\ \bibnamefont {Jaron-Becker}}, \bibinfo {author} {\bibfnamefont {Margaret~M.}\ \bibnamefont {Murnane}}, \ and\ \bibinfo {author} {\bibfnamefont {Henry~C.}\ \bibnamefont {Kapteyn}},\ }\bibfield  {title} {\enquote {\bibinfo {title} {Bright coherent ultrahigh harmonics in the kev x-ray regime from mid-infrared femtosecond lasers},}\ }\href {\doibase 10.1126/science.1218497} {\bibfield  {journal} {\bibinfo  {journal} {Science}\ }\textbf {\bibinfo {volume} {336}},\ \bibinfo {pages} {1287--1291} (\bibinfo {year} {2012})},\ \Eprint {http://arxiv.org/abs/https://www.science.org/doi/pdf/10.1126/science.1218497} {https://www.science.org/doi/pdf/10.1126/science.1218497} \BibitemShut {NoStop}%
\bibitem [{\citenamefont {Luu}\ \emph {et~al.}(2018)\citenamefont {Luu}, \citenamefont {Yin}, \citenamefont {Jain}, \citenamefont {Gaumnitz}, \citenamefont {Pertot}, \citenamefont {Ma},\ and\ \citenamefont {W{\"o}rner}}]{Luu2018}%
  \BibitemOpen
  \bibfield  {author} {\bibinfo {author} {\bibfnamefont {Tran~Trung}\ \bibnamefont {Luu}}, \bibinfo {author} {\bibfnamefont {Zhong}\ \bibnamefont {Yin}}, \bibinfo {author} {\bibfnamefont {Arohi}\ \bibnamefont {Jain}}, \bibinfo {author} {\bibfnamefont {Thomas}\ \bibnamefont {Gaumnitz}}, \bibinfo {author} {\bibfnamefont {Yoann}\ \bibnamefont {Pertot}}, \bibinfo {author} {\bibfnamefont {Jun}\ \bibnamefont {Ma}}, \ and\ \bibinfo {author} {\bibfnamefont {Hans~Jakob}\ \bibnamefont {W{\"o}rner}},\ }\bibfield  {title} {\enquote {\bibinfo {title} {Extreme--ultraviolet high--harmonic generation in liquids},}\ }\href {\doibase 10.1038/s41467-018-06040-4} {\bibfield  {journal} {\bibinfo  {journal} {Nature Communications}\ }\textbf {\bibinfo {volume} {9}},\ \bibinfo {pages} {3723} (\bibinfo {year} {2018})}\BibitemShut {NoStop}%
\bibitem [{\citenamefont {Mondal}\ \emph {et~al.}(2023)\citenamefont {Mondal}, \citenamefont {Waser}, \citenamefont {Balciunas}, \citenamefont {Neufeld}, \citenamefont {Yin}, \citenamefont {Tancogne-Dejean}, \citenamefont {Rubio},\ and\ \citenamefont {W\"{o}rner}}]{Mondal2023}%
  \BibitemOpen
  \bibfield  {author} {\bibinfo {author} {\bibfnamefont {Angana}\ \bibnamefont {Mondal}}, \bibinfo {author} {\bibfnamefont {Benedikt}\ \bibnamefont {Waser}}, \bibinfo {author} {\bibfnamefont {Tadas}\ \bibnamefont {Balciunas}}, \bibinfo {author} {\bibfnamefont {Ofer}\ \bibnamefont {Neufeld}}, \bibinfo {author} {\bibfnamefont {Zhong}\ \bibnamefont {Yin}}, \bibinfo {author} {\bibfnamefont {Nicolas}\ \bibnamefont {Tancogne-Dejean}}, \bibinfo {author} {\bibfnamefont {Angel}\ \bibnamefont {Rubio}}, \ and\ \bibinfo {author} {\bibfnamefont {Hans~Jakob}\ \bibnamefont {W\"{o}rner}},\ }\bibfield  {title} {\enquote {\bibinfo {title} {High-harmonic generation in liquids with few-cycle pulses: effect of laser-pulse duration on the cut-off energy},}\ }\href {\doibase 10.1364/OE.496686} {\bibfield  {journal} {\bibinfo  {journal} {Opt. Express}\ }\textbf {\bibinfo {volume} {31}},\ \bibinfo {pages} {34348--34361} (\bibinfo {year} {2023})}\BibitemShut {NoStop}%
\bibitem [{\citenamefont {Schubert}\ \emph {et~al.}(2014)\citenamefont {Schubert}, \citenamefont {Hohenleutner}, \citenamefont {Langer}, \citenamefont {Urbanek}, \citenamefont {Lange}, \citenamefont {Huttner}, \citenamefont {Golde}, \citenamefont {Meier}, \citenamefont {Kira}, \citenamefont {Koch},\ and\ \citenamefont {Huber}}]{Schubert2014}%
  \BibitemOpen
  \bibfield  {author} {\bibinfo {author} {\bibfnamefont {O.}~\bibnamefont {Schubert}}, \bibinfo {author} {\bibfnamefont {M.}~\bibnamefont {Hohenleutner}}, \bibinfo {author} {\bibfnamefont {F.}~\bibnamefont {Langer}}, \bibinfo {author} {\bibfnamefont {B.}~\bibnamefont {Urbanek}}, \bibinfo {author} {\bibfnamefont {C.}~\bibnamefont {Lange}}, \bibinfo {author} {\bibfnamefont {U.}~\bibnamefont {Huttner}}, \bibinfo {author} {\bibfnamefont {D.}~\bibnamefont {Golde}}, \bibinfo {author} {\bibfnamefont {T.}~\bibnamefont {Meier}}, \bibinfo {author} {\bibfnamefont {M.}~\bibnamefont {Kira}}, \bibinfo {author} {\bibfnamefont {S.~W.}\ \bibnamefont {Koch}}, \ and\ \bibinfo {author} {\bibfnamefont {R.}~\bibnamefont {Huber}},\ }\bibfield  {title} {\enquote {\bibinfo {title} {Sub-cycle control of terahertz high-harmonic generation by dynamical bloch oscillations},}\ }\href {\doibase 10.1038/nphoton.2013.349} {\bibfield  {journal} {\bibinfo  {journal} {Nature Photonics}\ }\textbf {\bibinfo {volume} {8}},\ \bibinfo {pages}
  {119--123} (\bibinfo {year} {2014})}\BibitemShut {NoStop}%
\bibitem [{\citenamefont {Ghimire}\ and\ \citenamefont {Reis}(2019)}]{Ghimire2019}%
  \BibitemOpen
  \bibfield  {author} {\bibinfo {author} {\bibfnamefont {Shambhu}\ \bibnamefont {Ghimire}}\ and\ \bibinfo {author} {\bibfnamefont {David~A.}\ \bibnamefont {Reis}},\ }\bibfield  {title} {\enquote {\bibinfo {title} {High-harmonic generation from solids},}\ }\href {\doibase 10.1038/s41567-018-0315-5} {\bibfield  {journal} {\bibinfo  {journal} {Nature Physics}\ }\textbf {\bibinfo {volume} {15}},\ \bibinfo {pages} {10--16} (\bibinfo {year} {2019})}\BibitemShut {NoStop}%
\bibitem [{\citenamefont {Ghimire}\ \emph {et~al.}(2011)\citenamefont {Ghimire}, \citenamefont {DiChiara}, \citenamefont {Sistrunk}, \citenamefont {Agostini}, \citenamefont {DiMauro},\ and\ \citenamefont {Reis}}]{Ghimire2011}%
  \BibitemOpen
  \bibfield  {author} {\bibinfo {author} {\bibfnamefont {Shambhu}\ \bibnamefont {Ghimire}}, \bibinfo {author} {\bibfnamefont {Anthony~D.}\ \bibnamefont {DiChiara}}, \bibinfo {author} {\bibfnamefont {Emily}\ \bibnamefont {Sistrunk}}, \bibinfo {author} {\bibfnamefont {Pierre}\ \bibnamefont {Agostini}}, \bibinfo {author} {\bibfnamefont {Louis~F.}\ \bibnamefont {DiMauro}}, \ and\ \bibinfo {author} {\bibfnamefont {David~A.}\ \bibnamefont {Reis}},\ }\bibfield  {title} {\enquote {\bibinfo {title} {Observation of high-order harmonic generation in a bulk crystal},}\ }\href {\doibase 10.1038/nphys1847} {\bibfield  {journal} {\bibinfo  {journal} {Nature Physics}\ }\textbf {\bibinfo {volume} {7}},\ \bibinfo {pages} {138--141} (\bibinfo {year} {2011})}\BibitemShut {NoStop}%
\bibitem [{\citenamefont {Wu}\ \emph {et~al.}(2015)\citenamefont {Wu}, \citenamefont {Ghimire}, \citenamefont {Reis}, \citenamefont {Schafer},\ and\ \citenamefont {Gaarde}}]{Wu2015}%
  \BibitemOpen
  \bibfield  {author} {\bibinfo {author} {\bibfnamefont {Mengxi}\ \bibnamefont {Wu}}, \bibinfo {author} {\bibfnamefont {Shambhu}\ \bibnamefont {Ghimire}}, \bibinfo {author} {\bibfnamefont {David~A.}\ \bibnamefont {Reis}}, \bibinfo {author} {\bibfnamefont {Kenneth~J.}\ \bibnamefont {Schafer}}, \ and\ \bibinfo {author} {\bibfnamefont {Mette~B.}\ \bibnamefont {Gaarde}},\ }\bibfield  {title} {\enquote {\bibinfo {title} {High-harmonic generation from bloch electrons in solids},}\ }\href {\doibase 10.1103/PhysRevA.91.043839} {\bibfield  {journal} {\bibinfo  {journal} {Phys. Rev. A}\ }\textbf {\bibinfo {volume} {91}},\ \bibinfo {pages} {043839} (\bibinfo {year} {2015})}\BibitemShut {NoStop}%
\bibitem [{\citenamefont {Corkum}\ and\ \citenamefont {Krausz}(2007)}]{Corkum2007}%
  \BibitemOpen
  \bibfield  {author} {\bibinfo {author} {\bibfnamefont {P.~B.}\ \bibnamefont {Corkum}}\ and\ \bibinfo {author} {\bibfnamefont {Ferenc}\ \bibnamefont {Krausz}},\ }\bibfield  {title} {\enquote {\bibinfo {title} {Attosecond science},}\ }\href {\doibase 10.1038/nphys620} {\bibfield  {journal} {\bibinfo  {journal} {Nature Physics}\ }\textbf {\bibinfo {volume} {3}},\ \bibinfo {pages} {381--387} (\bibinfo {year} {2007})}\BibitemShut {NoStop}%
\bibitem [{\citenamefont {Ghimire}\ \emph {et~al.}(2014)\citenamefont {Ghimire}, \citenamefont {Ndabashimiye}, \citenamefont {DiChiara}, \citenamefont {Sistrunk}, \citenamefont {Stockman}, \citenamefont {Agostini}, \citenamefont {DiMauro},\ and\ \citenamefont {Reis}}]{Ghimire2014}%
  \BibitemOpen
  \bibfield  {author} {\bibinfo {author} {\bibfnamefont {Shambhu}\ \bibnamefont {Ghimire}}, \bibinfo {author} {\bibfnamefont {Georges}\ \bibnamefont {Ndabashimiye}}, \bibinfo {author} {\bibfnamefont {Anthony~D}\ \bibnamefont {DiChiara}}, \bibinfo {author} {\bibfnamefont {Emily}\ \bibnamefont {Sistrunk}}, \bibinfo {author} {\bibfnamefont {Mark~I}\ \bibnamefont {Stockman}}, \bibinfo {author} {\bibfnamefont {Pierre}\ \bibnamefont {Agostini}}, \bibinfo {author} {\bibfnamefont {Louis~F}\ \bibnamefont {DiMauro}}, \ and\ \bibinfo {author} {\bibfnamefont {David~A}\ \bibnamefont {Reis}},\ }\bibfield  {title} {\enquote {\bibinfo {title} {Strong-field and attosecond physics in solids},}\ }\href {\doibase 10.1088/0953-4075/47/20/204030} {\bibfield  {journal} {\bibinfo  {journal} {Journal of Physics B: Atomic, Molecular and Optical Physics}\ }\textbf {\bibinfo {volume} {47}},\ \bibinfo {pages} {204030} (\bibinfo {year} {2014})}\BibitemShut {NoStop}%
\bibitem [{\citenamefont {Krausz}(2016)}]{Krausz2016}%
  \BibitemOpen
  \bibfield  {author} {\bibinfo {author} {\bibfnamefont {Ferenc}\ \bibnamefont {Krausz}},\ }\bibfield  {title} {\enquote {\bibinfo {title} {The birth of attosecond physics and its coming of age},}\ }\href {\doibase 10.1088/0031-8949/91/6/063011} {\bibfield  {journal} {\bibinfo  {journal} {Physica Scripta}\ }\textbf {\bibinfo {volume} {91}},\ \bibinfo {pages} {063011} (\bibinfo {year} {2016})}\BibitemShut {NoStop}%
\bibitem [{\citenamefont {Krausz}\ and\ \citenamefont {Ivanov}(2009)}]{Krausz}%
  \BibitemOpen
  \bibfield  {author} {\bibinfo {author} {\bibfnamefont {Ferenc}\ \bibnamefont {Krausz}}\ and\ \bibinfo {author} {\bibfnamefont {Misha}\ \bibnamefont {Ivanov}},\ }\bibfield  {title} {\enquote {\bibinfo {title} {Attosecond physics},}\ }\href {\doibase 10.1103/RevModPhys.81.163} {\bibfield  {journal} {\bibinfo  {journal} {Rev. Mod. Phys.}\ }\textbf {\bibinfo {volume} {81}},\ \bibinfo {pages} {163--234} (\bibinfo {year} {2009})}\BibitemShut {NoStop}%
\bibitem [{\citenamefont {Ly}\ and\ \citenamefont {Manchon}(2022)}]{Ly2022}%
  \BibitemOpen
  \bibfield  {author} {\bibinfo {author} {\bibfnamefont {Ousmane}\ \bibnamefont {Ly}}\ and\ \bibinfo {author} {\bibfnamefont {Aurelien}\ \bibnamefont {Manchon}},\ }\bibfield  {title} {\enquote {\bibinfo {title} {Spin-orbit coupling induced ultrahigh-harmonic generation from magnetic dynamics},}\ }\href {\doibase 10.1103/PhysRevB.105.L180415} {\bibfield  {journal} {\bibinfo  {journal} {Phys. Rev. B}\ }\textbf {\bibinfo {volume} {105}},\ \bibinfo {pages} {L180415} (\bibinfo {year} {2022})}\BibitemShut {NoStop}%
\bibitem [{\citenamefont {Ly}(2023{\natexlab{a}})}]{Ly2023}%
  \BibitemOpen
  \bibfield  {author} {\bibinfo {author} {\bibfnamefont {Ousmane}\ \bibnamefont {Ly}},\ }\bibfield  {title} {\enquote {\bibinfo {title} {Noncollinear antiferromagnetic textures driven high-harmonic generation from magnetic dynamics in the absence of spin-orbit coupling},}\ }\href {\doibase 10.1088/1361-648X/acb523} {\bibfield  {journal} {\bibinfo  {journal} {Journal of Physics: Condensed Matter}\ }\textbf {\bibinfo {volume} {35}},\ \bibinfo {pages} {125802} (\bibinfo {year} {2023}{\natexlab{a}})}\BibitemShut {NoStop}%
\bibitem [{\citenamefont {Tserkovnyak}\ \emph {et~al.}(2002{\natexlab{a}})\citenamefont {Tserkovnyak}, \citenamefont {Brataas},\ and\ \citenamefont {Bauer}}]{Tserkovnyak2002}%
  \BibitemOpen
  \bibfield  {author} {\bibinfo {author} {\bibfnamefont {Yaroslav}\ \bibnamefont {Tserkovnyak}}, \bibinfo {author} {\bibfnamefont {Arne}\ \bibnamefont {Brataas}}, \ and\ \bibinfo {author} {\bibfnamefont {Gerrit E.~W.}\ \bibnamefont {Bauer}},\ }\bibfield  {title} {\enquote {\bibinfo {title} {Enhanced gilbert damping in thin ferromagnetic films},}\ }\href {\doibase 10.1103/PhysRevLett.88.117601} {\bibfield  {journal} {\bibinfo  {journal} {Phys. Rev. Lett.}\ }\textbf {\bibinfo {volume} {88}},\ \bibinfo {pages} {117601} (\bibinfo {year} {2002}{\natexlab{a}})}\BibitemShut {NoStop}%
\bibitem [{\citenamefont {Tserkovnyak}\ \emph {et~al.}(2002{\natexlab{b}})\citenamefont {Tserkovnyak}, \citenamefont {Brataas},\ and\ \citenamefont {Bauer}}]{Tserkovnyak2002b}%
  \BibitemOpen
  \bibfield  {author} {\bibinfo {author} {\bibfnamefont {Yaroslav}\ \bibnamefont {Tserkovnyak}}, \bibinfo {author} {\bibfnamefont {Arne}\ \bibnamefont {Brataas}}, \ and\ \bibinfo {author} {\bibfnamefont {Gerrit E.~W.}\ \bibnamefont {Bauer}},\ }\bibfield  {title} {\enquote {\bibinfo {title} {Spin pumping and magnetization dynamics in metallic multilayers},}\ }\href {\doibase 10.1103/PhysRevB.66.224403} {\bibfield  {journal} {\bibinfo  {journal} {Phys. Rev. B}\ }\textbf {\bibinfo {volume} {66}},\ \bibinfo {pages} {224403} (\bibinfo {year} {2002}{\natexlab{b}})}\BibitemShut {NoStop}%
\bibitem [{\citenamefont {Tserkovnyak}\ \emph {et~al.}(2005)\citenamefont {Tserkovnyak}, \citenamefont {Brataas}, \citenamefont {Bauer},\ and\ \citenamefont {Halperin}}]{Tserkovnyak2005}%
  \BibitemOpen
  \bibfield  {author} {\bibinfo {author} {\bibfnamefont {Yaroslav}\ \bibnamefont {Tserkovnyak}}, \bibinfo {author} {\bibfnamefont {Arne}\ \bibnamefont {Brataas}}, \bibinfo {author} {\bibfnamefont {Gerrit E.~W.}\ \bibnamefont {Bauer}}, \ and\ \bibinfo {author} {\bibfnamefont {Bertrand~I.}\ \bibnamefont {Halperin}},\ }\bibfield  {title} {\enquote {\bibinfo {title} {Nonlocal magnetization dynamics in ferromagnetic heterostructures},}\ }\href {\doibase 10.1103/RevModPhys.77.1375} {\bibfield  {journal} {\bibinfo  {journal} {Rev. Mod. Phys.}\ }\textbf {\bibinfo {volume} {77}},\ \bibinfo {pages} {1375--1421} (\bibinfo {year} {2005})}\BibitemShut {NoStop}%
\bibitem [{\citenamefont {Cheng}\ \emph {et~al.}(2014)\citenamefont {Cheng}, \citenamefont {Xiao}, \citenamefont {Niu},\ and\ \citenamefont {Brataas}}]{Cheng2014}%
  \BibitemOpen
  \bibfield  {author} {\bibinfo {author} {\bibfnamefont {Ran}\ \bibnamefont {Cheng}}, \bibinfo {author} {\bibfnamefont {Jiang}\ \bibnamefont {Xiao}}, \bibinfo {author} {\bibfnamefont {Qian}\ \bibnamefont {Niu}}, \ and\ \bibinfo {author} {\bibfnamefont {Arne}\ \bibnamefont {Brataas}},\ }\bibfield  {title} {\enquote {\bibinfo {title} {Spin pumping and spin-transfer torques in antiferromagnets},}\ }\href {\doibase 10.1103/PhysRevLett.113.057601} {\bibfield  {journal} {\bibinfo  {journal} {Phys. Rev. Lett.}\ }\textbf {\bibinfo {volume} {113}},\ \bibinfo {pages} {057601} (\bibinfo {year} {2014})}\BibitemShut {NoStop}%
\bibitem [{\citenamefont {Vaidya}\ \emph {et~al.}(2020)\citenamefont {Vaidya}, \citenamefont {Morley}, \citenamefont {van Tol}, \citenamefont {Liu}, \citenamefont {Cheng}, \citenamefont {Brataas}, \citenamefont {Lederman},\ and\ \citenamefont {del Barco}}]{Vaidya2020}%
  \BibitemOpen
  \bibfield  {author} {\bibinfo {author} {\bibfnamefont {Priyanka}\ \bibnamefont {Vaidya}}, \bibinfo {author} {\bibfnamefont {Sophie~A.}\ \bibnamefont {Morley}}, \bibinfo {author} {\bibfnamefont {Johan}\ \bibnamefont {van Tol}}, \bibinfo {author} {\bibfnamefont {Yan}\ \bibnamefont {Liu}}, \bibinfo {author} {\bibfnamefont {Ran}\ \bibnamefont {Cheng}}, \bibinfo {author} {\bibfnamefont {Arne}\ \bibnamefont {Brataas}}, \bibinfo {author} {\bibfnamefont {David}\ \bibnamefont {Lederman}}, \ and\ \bibinfo {author} {\bibfnamefont {Enrique}\ \bibnamefont {del Barco}},\ }\bibfield  {title} {\enquote {\bibinfo {title} {Subterahertz spin pumping from an insulating antiferromagnet},}\ }\href {\doibase 10.1126/science.aaz4247} {\bibfield  {journal} {\bibinfo  {journal} {Science}\ }\textbf {\bibinfo {volume} {368}},\ \bibinfo {pages} {160--165} (\bibinfo {year} {2020})},\ \Eprint {http://arxiv.org/abs/https://www.science.org/doi/pdf/10.1126/science.aaz4247} {https://www.science.org/doi/pdf/10.1126/science.aaz4247}
  \BibitemShut {NoStop}%
\bibitem [{\citenamefont {St\ifmmode~\check{r}\else \v{r}\fi{}eda}\ and\ \citenamefont {\ifmmode~\check{S}\else \v{S}\fi{}eba}(2003)}]{Streda2003}%
  \BibitemOpen
  \bibfield  {author} {\bibinfo {author} {\bibfnamefont {P.}~\bibnamefont {St\ifmmode~\check{r}\else \v{r}\fi{}eda}}\ and\ \bibinfo {author} {\bibfnamefont {P.}~\bibnamefont {\ifmmode~\check{S}\else \v{S}\fi{}eba}},\ }\bibfield  {title} {\enquote {\bibinfo {title} {Antisymmetric spin filtering in one-dimensional electron systems with uniform spin-orbit coupling},}\ }\href {\doibase 10.1103/PhysRevLett.90.256601} {\bibfield  {journal} {\bibinfo  {journal} {Phys. Rev. Lett.}\ }\textbf {\bibinfo {volume} {90}},\ \bibinfo {pages} {256601} (\bibinfo {year} {2003})}\BibitemShut {NoStop}%
\bibitem [{\citenamefont {Birkholz}\ and\ \citenamefont {Meden}(2008)}]{Birkholz2008}%
  \BibitemOpen
  \bibfield  {author} {\bibinfo {author} {\bibfnamefont {J~E}\ \bibnamefont {Birkholz}}\ and\ \bibinfo {author} {\bibfnamefont {V}~\bibnamefont {Meden}},\ }\bibfield  {title} {\enquote {\bibinfo {title} {Spin–orbit coupling effects in one-dimensional ballistic quantum wires},}\ }\href {\doibase 10.1088/0953-8984/20/8/085226} {\bibfield  {journal} {\bibinfo  {journal} {Journal of Physics: Condensed Matter}\ }\textbf {\bibinfo {volume} {20}},\ \bibinfo {pages} {085226} (\bibinfo {year} {2008})}\BibitemShut {NoStop}%
\bibitem [{\citenamefont {Kloss}\ \emph {et~al.}(2021)\citenamefont {Kloss}, \citenamefont {Weston}, \citenamefont {Gaury}, \citenamefont {Rossignol}, \citenamefont {Groth},\ and\ \citenamefont {Waintal}}]{tkwant}%
  \BibitemOpen
  \bibfield  {author} {\bibinfo {author} {\bibfnamefont {Thomas}\ \bibnamefont {Kloss}}, \bibinfo {author} {\bibfnamefont {Joseph}\ \bibnamefont {Weston}}, \bibinfo {author} {\bibfnamefont {Benoit}\ \bibnamefont {Gaury}}, \bibinfo {author} {\bibfnamefont {Benoit}\ \bibnamefont {Rossignol}}, \bibinfo {author} {\bibfnamefont {Christoph}\ \bibnamefont {Groth}}, \ and\ \bibinfo {author} {\bibfnamefont {Xavier}\ \bibnamefont {Waintal}},\ }\bibfield  {title} {\enquote {\bibinfo {title} {Tkwant: a software package for time-dependent quantum transport},}\ }\href {\doibase 10.1088/1367-2630/abddf7} {\bibfield  {journal} {\bibinfo  {journal} {New Journal of Physics}\ }\textbf {\bibinfo {volume} {23}},\ \bibinfo {pages} {023025} (\bibinfo {year} {2021})}\BibitemShut {NoStop}%
\bibitem [{\citenamefont {Xiao}\ \emph {et~al.}(2010)\citenamefont {Xiao}, \citenamefont {Chang},\ and\ \citenamefont {Niu}}]{Xiao2010}%
  \BibitemOpen
  \bibfield  {author} {\bibinfo {author} {\bibfnamefont {Di}~\bibnamefont {Xiao}}, \bibinfo {author} {\bibfnamefont {Ming-Che}\ \bibnamefont {Chang}}, \ and\ \bibinfo {author} {\bibfnamefont {Qian}\ \bibnamefont {Niu}},\ }\bibfield  {title} {\enquote {\bibinfo {title} {Berry phase effects on electronic properties},}\ }\href {\doibase 10.1103/RevModPhys.82.1959} {\bibfield  {journal} {\bibinfo  {journal} {Rev. Mod. Phys.}\ }\textbf {\bibinfo {volume} {82}},\ \bibinfo {pages} {1959--2007} (\bibinfo {year} {2010})}\BibitemShut {NoStop}%
\bibitem [{\citenamefont {Ly}(2023{\natexlab{b}})}]{Ly2023c}%
  \BibitemOpen
  \bibfield  {author} {\bibinfo {author} {\bibfnamefont {O.}~\bibnamefont {Ly}},\ }\bibfield  {title} {\enquote {\bibinfo {title} {{Scaling laws of the magnetic dynamics driven high harmonic generation in spin-orbit coupled systems}},}\ }\href@noop {} {\  (\bibinfo {year} {2023}{\natexlab{b}})},\ \Eprint {http://arxiv.org/abs/arXiv:2304.02619} {arXiv:arXiv:2304.02619} \BibitemShut {NoStop}%
\bibitem [{\citenamefont {Kiselev}\ \emph {et~al.}(2003)\citenamefont {Kiselev}, \citenamefont {Sankey}, \citenamefont {Krivorotov}, \citenamefont {Emley}, \citenamefont {Schoelkopf}, \citenamefont {Buhrman},\ and\ \citenamefont {Ralph}}]{Kiselev2003}%
  \BibitemOpen
  \bibfield  {author} {\bibinfo {author} {\bibfnamefont {S.~I.}\ \bibnamefont {Kiselev}}, \bibinfo {author} {\bibfnamefont {J.~C.}\ \bibnamefont {Sankey}}, \bibinfo {author} {\bibfnamefont {I.~N.}\ \bibnamefont {Krivorotov}}, \bibinfo {author} {\bibfnamefont {N.~C.}\ \bibnamefont {Emley}}, \bibinfo {author} {\bibfnamefont {R.~J.}\ \bibnamefont {Schoelkopf}}, \bibinfo {author} {\bibfnamefont {R.~A.}\ \bibnamefont {Buhrman}}, \ and\ \bibinfo {author} {\bibfnamefont {D.~C.}\ \bibnamefont {Ralph}},\ }\bibfield  {title} {\enquote {\bibinfo {title} {Microwave oscillations of a nanomagnet driven by a spin-polarized current},}\ }\href {\doibase 10.1038/nature01967} {\bibfield  {journal} {\bibinfo  {journal} {Nature}\ }\textbf {\bibinfo {volume} {425}},\ \bibinfo {pages} {380--383} (\bibinfo {year} {2003})}\BibitemShut {NoStop}%
\bibitem [{\citenamefont {Liu}\ \emph {et~al.}(2012)\citenamefont {Liu}, \citenamefont {Pai}, \citenamefont {Ralph},\ and\ \citenamefont {Buhrman}}]{Liu2012}%
  \BibitemOpen
  \bibfield  {author} {\bibinfo {author} {\bibfnamefont {Luqiao}\ \bibnamefont {Liu}}, \bibinfo {author} {\bibfnamefont {Chi-Feng}\ \bibnamefont {Pai}}, \bibinfo {author} {\bibfnamefont {D.~C.}\ \bibnamefont {Ralph}}, \ and\ \bibinfo {author} {\bibfnamefont {R.~A.}\ \bibnamefont {Buhrman}},\ }\bibfield  {title} {\enquote {\bibinfo {title} {Magnetic oscillations driven by the spin hall effect in 3-terminal magnetic tunnel junction devices},}\ }\href {\doibase 10.1103/PhysRevLett.109.186602} {\bibfield  {journal} {\bibinfo  {journal} {Phys. Rev. Lett.}\ }\textbf {\bibinfo {volume} {109}},\ \bibinfo {pages} {186602} (\bibinfo {year} {2012})}\BibitemShut {NoStop}%
\bibitem [{\citenamefont {Cooper}\ and\ \citenamefont {Uehling}(1967)}]{Cooper1967}%
  \BibitemOpen
  \bibfield  {author} {\bibinfo {author} {\bibfnamefont {Robert~L.}\ \bibnamefont {Cooper}}\ and\ \bibinfo {author} {\bibfnamefont {Edwin~A.}\ \bibnamefont {Uehling}},\ }\bibfield  {title} {\enquote {\bibinfo {title} {Ferromagnetic resonance and spin diffusion in supermalloy},}\ }\href {\doibase 10.1103/PhysRev.164.662} {\bibfield  {journal} {\bibinfo  {journal} {Phys. Rev.}\ }\textbf {\bibinfo {volume} {164}},\ \bibinfo {pages} {662--668} (\bibinfo {year} {1967})}\BibitemShut {NoStop}%
\bibitem [{\citenamefont {Xiao}\ \emph {et~al.}(2008)\citenamefont {Xiao}, \citenamefont {Bauer},\ and\ \citenamefont {Brataas}}]{Xiao2008}%
  \BibitemOpen
  \bibfield  {author} {\bibinfo {author} {\bibfnamefont {Jiang}\ \bibnamefont {Xiao}}, \bibinfo {author} {\bibfnamefont {Gerrit E.~W.}\ \bibnamefont {Bauer}}, \ and\ \bibinfo {author} {\bibfnamefont {Arne}\ \bibnamefont {Brataas}},\ }\bibfield  {title} {\enquote {\bibinfo {title} {Charge pumping in magnetic tunnel junctions: Scattering theory},}\ }\href {\doibase 10.1103/PhysRevB.77.180407} {\bibfield  {journal} {\bibinfo  {journal} {Phys. Rev. B}\ }\textbf {\bibinfo {volume} {77}},\ \bibinfo {pages} {180407} (\bibinfo {year} {2008})}\BibitemShut {NoStop}%
\bibitem [{\citenamefont {Ast}\ \emph {et~al.}(2007)\citenamefont {Ast}, \citenamefont {Henk}, \citenamefont {Ernst}, \citenamefont {Moreschini}, \citenamefont {Falub}, \citenamefont {Pacil\'e}, \citenamefont {Bruno}, \citenamefont {Kern},\ and\ \citenamefont {Grioni}}]{Ast2007}%
  \BibitemOpen
  \bibfield  {author} {\bibinfo {author} {\bibfnamefont {Christian~R.}\ \bibnamefont {Ast}}, \bibinfo {author} {\bibfnamefont {J\"urgen}\ \bibnamefont {Henk}}, \bibinfo {author} {\bibfnamefont {Arthur}\ \bibnamefont {Ernst}}, \bibinfo {author} {\bibfnamefont {Luca}\ \bibnamefont {Moreschini}}, \bibinfo {author} {\bibfnamefont {Mihaela~C.}\ \bibnamefont {Falub}}, \bibinfo {author} {\bibfnamefont {Daniela}\ \bibnamefont {Pacil\'e}}, \bibinfo {author} {\bibfnamefont {Patrick}\ \bibnamefont {Bruno}}, \bibinfo {author} {\bibfnamefont {Klaus}\ \bibnamefont {Kern}}, \ and\ \bibinfo {author} {\bibfnamefont {Marco}\ \bibnamefont {Grioni}},\ }\bibfield  {title} {\enquote {\bibinfo {title} {Giant spin splitting through surface alloying},}\ }\href {\doibase 10.1103/PhysRevLett.98.186807} {\bibfield  {journal} {\bibinfo  {journal} {Phys. Rev. Lett.}\ }\textbf {\bibinfo {volume} {98}},\ \bibinfo {pages} {186807} (\bibinfo {year} {2007})}\BibitemShut {NoStop}%
\bibitem [{\citenamefont {Zhang}\ \emph {et~al.}(2009)\citenamefont {Zhang}, \citenamefont {Liu}, \citenamefont {Qi}, \citenamefont {Dai}, \citenamefont {Fang},\ and\ \citenamefont {Zhang}}]{Zhang2009}%
  \BibitemOpen
  \bibfield  {author} {\bibinfo {author} {\bibfnamefont {Haijun}\ \bibnamefont {Zhang}}, \bibinfo {author} {\bibfnamefont {Chao-Xing}\ \bibnamefont {Liu}}, \bibinfo {author} {\bibfnamefont {Xiao-Liang}\ \bibnamefont {Qi}}, \bibinfo {author} {\bibfnamefont {Xi}~\bibnamefont {Dai}}, \bibinfo {author} {\bibfnamefont {Zhong}\ \bibnamefont {Fang}}, \ and\ \bibinfo {author} {\bibfnamefont {Shou-Cheng}\ \bibnamefont {Zhang}},\ }\bibfield  {title} {\enquote {\bibinfo {title} {Topological insulators in bi2se3, bi2te3 and sb2te3 with a single dirac cone on the surface},}\ }\href {\doibase 10.1038/nphys1270} {\bibfield  {journal} {\bibinfo  {journal} {Nature Physics}\ }\textbf {\bibinfo {volume} {5}},\ \bibinfo {pages} {438--442} (\bibinfo {year} {2009})}\BibitemShut {NoStop}%
\bibitem [{\citenamefont {Neufeld}\ \emph {et~al.}(2022)\citenamefont {Neufeld}, \citenamefont {Mao}, \citenamefont {H\"ubener}, \citenamefont {Tancogne-Dejean}, \citenamefont {Sato}, \citenamefont {De~Giovannini},\ and\ \citenamefont {Rubio}}]{Neufeld2022}%
  \BibitemOpen
  \bibfield  {author} {\bibinfo {author} {\bibfnamefont {Ofer}\ \bibnamefont {Neufeld}}, \bibinfo {author} {\bibfnamefont {Wenwen}\ \bibnamefont {Mao}}, \bibinfo {author} {\bibfnamefont {Hannes}\ \bibnamefont {H\"ubener}}, \bibinfo {author} {\bibfnamefont {Nicolas}\ \bibnamefont {Tancogne-Dejean}}, \bibinfo {author} {\bibfnamefont {Shunsuke~A.}\ \bibnamefont {Sato}}, \bibinfo {author} {\bibfnamefont {Umberto}\ \bibnamefont {De~Giovannini}}, \ and\ \bibinfo {author} {\bibfnamefont {Angel}\ \bibnamefont {Rubio}},\ }\bibfield  {title} {\enquote {\bibinfo {title} {Time- and angle-resolved photoelectron spectroscopy of strong-field light-dressed solids: Prevalence of the adiabatic band picture},}\ }\href {\doibase 10.1103/PhysRevResearch.4.033101} {\bibfield  {journal} {\bibinfo  {journal} {Phys. Rev. Res.}\ }\textbf {\bibinfo {volume} {4}},\ \bibinfo {pages} {033101} (\bibinfo {year} {2022})}\BibitemShut {NoStop}%
\bibitem [{\citenamefont {Gürsoy}\ \emph {et~al.}(2023)\citenamefont {Gürsoy}, \citenamefont {Reck}, \citenamefont {Gorini}, \citenamefont {Richter},\ and\ \citenamefont {Adagideli}}]{Gursoy2023}%
  \BibitemOpen
  \bibfield  {author} {\bibinfo {author} {\bibfnamefont {Fahriye~N.}\ \bibnamefont {Gürsoy}}, \bibinfo {author} {\bibfnamefont {P.}~\bibnamefont {Reck}}, \bibinfo {author} {\bibfnamefont {C.}~\bibnamefont {Gorini}}, \bibinfo {author} {\bibfnamefont {K.}~\bibnamefont {Richter}}, \ and\ \bibinfo {author} {\bibfnamefont {I.}~\bibnamefont {Adagideli}},\ }\bibfield  {title} {\enquote {\bibinfo {title} {{Dynamical spin-orbit-based spin transistor}},}\ }\href {\doibase 10.21468/SciPostPhys.14.4.060} {\bibfield  {journal} {\bibinfo  {journal} {SciPost Phys.}\ }\textbf {\bibinfo {volume} {14}},\ \bibinfo {pages} {060} (\bibinfo {year} {2023})}\BibitemShut {NoStop}%
\bibitem [{\citenamefont {Imai}\ \emph {et~al.}(2020)\citenamefont {Imai}, \citenamefont {Ono},\ and\ \citenamefont {Ishihara}}]{Imai2020}%
  \BibitemOpen
  \bibfield  {author} {\bibinfo {author} {\bibfnamefont {Shohei}\ \bibnamefont {Imai}}, \bibinfo {author} {\bibfnamefont {Atsushi}\ \bibnamefont {Ono}}, \ and\ \bibinfo {author} {\bibfnamefont {Sumio}\ \bibnamefont {Ishihara}},\ }\bibfield  {title} {\enquote {\bibinfo {title} {High harmonic generation in a correlated electron system},}\ }\href {\doibase 10.1103/PhysRevLett.124.157404} {\bibfield  {journal} {\bibinfo  {journal} {Phys. Rev. Lett.}\ }\textbf {\bibinfo {volume} {124}},\ \bibinfo {pages} {157404} (\bibinfo {year} {2020})}\BibitemShut {NoStop}%
\bibitem [{\citenamefont {Hadadi}\ \emph {et~al.}(2023)\citenamefont {Hadadi}, \citenamefont {Belayadi}, \citenamefont {AlRabiah}, \citenamefont {Ly}, \citenamefont {Akosa}, \citenamefont {Vogl}, \citenamefont {Bahlouli}, \citenamefont {Manchon},\ and\ \citenamefont {Abbout}}]{Hadadi2023}%
  \BibitemOpen
  \bibfield  {author} {\bibinfo {author} {\bibfnamefont {Naif}\ \bibnamefont {Hadadi}}, \bibinfo {author} {\bibfnamefont {Adel}\ \bibnamefont {Belayadi}}, \bibinfo {author} {\bibfnamefont {Ahmed}\ \bibnamefont {AlRabiah}}, \bibinfo {author} {\bibfnamefont {Ousmane}\ \bibnamefont {Ly}}, \bibinfo {author} {\bibfnamefont {Collins~Ashu}\ \bibnamefont {Akosa}}, \bibinfo {author} {\bibfnamefont {Michael}\ \bibnamefont {Vogl}}, \bibinfo {author} {\bibfnamefont {Hocine}\ \bibnamefont {Bahlouli}}, \bibinfo {author} {\bibfnamefont {Aurelien}\ \bibnamefont {Manchon}}, \ and\ \bibinfo {author} {\bibfnamefont {Adel}\ \bibnamefont {Abbout}},\ }\bibfield  {title} {\enquote {\bibinfo {title} {Pseudo electric field and pumping valley current in graphene nanobubbles},}\ }\href {\doibase 10.1103/PhysRevB.108.195418} {\bibfield  {journal} {\bibinfo  {journal} {Phys. Rev. B}\ }\textbf {\bibinfo {volume} {108}},\ \bibinfo {pages} {195418} (\bibinfo {year} {2023})}\BibitemShut {NoStop}%
\bibitem [{\citenamefont {Schmid}\ \emph {et~al.}(2021)\citenamefont {Schmid}, \citenamefont {Weigl}, \citenamefont {Gr{\"o}ssing}, \citenamefont {Junk}, \citenamefont {Gorini}, \citenamefont {Schlauderer}, \citenamefont {Ito}, \citenamefont {Meierhofer}, \citenamefont {Hofmann}, \citenamefont {Afanasiev}, \citenamefont {Crewse}, \citenamefont {Kokh}, \citenamefont {Tereshchenko}, \citenamefont {G{\"u}dde}, \citenamefont {Evers}, \citenamefont {Wilhelm}, \citenamefont {Richter}, \citenamefont {H{\"o}fer},\ and\ \citenamefont {Huber}}]{Schmid2021}%
  \BibitemOpen
  \bibfield  {author} {\bibinfo {author} {\bibfnamefont {C.~P.}\ \bibnamefont {Schmid}}, \bibinfo {author} {\bibfnamefont {L.}~\bibnamefont {Weigl}}, \bibinfo {author} {\bibfnamefont {P.}~\bibnamefont {Gr{\"o}ssing}}, \bibinfo {author} {\bibfnamefont {V.}~\bibnamefont {Junk}}, \bibinfo {author} {\bibfnamefont {C.}~\bibnamefont {Gorini}}, \bibinfo {author} {\bibfnamefont {S.}~\bibnamefont {Schlauderer}}, \bibinfo {author} {\bibfnamefont {S.}~\bibnamefont {Ito}}, \bibinfo {author} {\bibfnamefont {M.}~\bibnamefont {Meierhofer}}, \bibinfo {author} {\bibfnamefont {N.}~\bibnamefont {Hofmann}}, \bibinfo {author} {\bibfnamefont {D.}~\bibnamefont {Afanasiev}}, \bibinfo {author} {\bibfnamefont {J.}~\bibnamefont {Crewse}}, \bibinfo {author} {\bibfnamefont {K.~A.}\ \bibnamefont {Kokh}}, \bibinfo {author} {\bibfnamefont {O.~E.}\ \bibnamefont {Tereshchenko}}, \bibinfo {author} {\bibfnamefont {J.}~\bibnamefont {G{\"u}dde}}, \bibinfo {author} {\bibfnamefont {F.}~\bibnamefont {Evers}}, \bibinfo {author} {\bibfnamefont
  {J.}~\bibnamefont {Wilhelm}}, \bibinfo {author} {\bibfnamefont {K.}~\bibnamefont {Richter}}, \bibinfo {author} {\bibfnamefont {U.}~\bibnamefont {H{\"o}fer}}, \ and\ \bibinfo {author} {\bibfnamefont {R.}~\bibnamefont {Huber}},\ }\bibfield  {title} {\enquote {\bibinfo {title} {Tunable non-integer high-harmonic generation in a topological insulator},}\ }\href {\doibase 10.1038/s41586-021-03466-7} {\bibfield  {journal} {\bibinfo  {journal} {Nature}\ }\textbf {\bibinfo {volume} {593}},\ \bibinfo {pages} {385--390} (\bibinfo {year} {2021})}\BibitemShut {NoStop}%
\bibitem [{\citenamefont {Lysne}\ \emph {et~al.}(2020)\citenamefont {Lysne}, \citenamefont {Murakami}, \citenamefont {Sch\"uler},\ and\ \citenamefont {Werner}}]{Lysne2020}%
  \BibitemOpen
  \bibfield  {author} {\bibinfo {author} {\bibfnamefont {Markus}\ \bibnamefont {Lysne}}, \bibinfo {author} {\bibfnamefont {Yuta}\ \bibnamefont {Murakami}}, \bibinfo {author} {\bibfnamefont {Michael}\ \bibnamefont {Sch\"uler}}, \ and\ \bibinfo {author} {\bibfnamefont {Philipp}\ \bibnamefont {Werner}},\ }\bibfield  {title} {\enquote {\bibinfo {title} {High-harmonic generation in spin-orbit coupled systems},}\ }\href {\doibase 10.1103/PhysRevB.102.081121} {\bibfield  {journal} {\bibinfo  {journal} {Phys. Rev. B}\ }\textbf {\bibinfo {volume} {102}},\ \bibinfo {pages} {081121} (\bibinfo {year} {2020})}\BibitemShut {NoStop}%
\end{thebibliography}%
\end{document}